\documentstyle[12pt]{article}
\begin{document}
\begin{flushright}
SINP/TNP/94-9
\end{flushright}
\bigskip
\begin{center}
{\Large\bf Baryogenesis at the Electroweak Scale}\\[1 cm]
\end{center}
\begin{center}
A. Kundu and S. Mallik\\
Saha Institute of Nuclear Physics\\
1/AF, Bidhannagar, Calcutta 700064, India
\end{center}
\vspace{1 cm}
\begin{abstract}
The generation of the baryon asymmetry of the universe is considered in
the standard model of the electroweak theory with simple extensions of the
Higgs sector. The propagation of quarks of masses up to about 5 GeV are
considered, taking into account their markedly different dispersion
relations due to propagation through the hot electroweak plasma. It is shown
that the contribution of these lighter quarks to the baryon asymmetry can
be comparable to that for the t quark considered earlier.
\end{abstract}
\newpage
\noindent{\bf I. INTRODUCTION}

\bigskip

It has been an extremely interesting observation that all the conditions
needed for baryogenesis [1] are already
present, in principle, in the
standard electroweak theory. The baryon number violation in this theory,
although exceedingly suppressed at the present time [2], can be unsuppressed
at high temperature [3]. C and CP violations are contained in the interaction
of quarks with Higgs fields. Finally departure from thermal equilibrium,
although difficult to obtain at the electroweak scale -- typical weak
interaction rates are extremely faster than the expansion rate of the
Universe -- can nevertheless exist if the electroweak phase transition is of
first order. We thus have the exciting possibility of explaining one of the
most fundamental problems of cosmology in terms of laboratory physics.

Several mechanisms have been proposed to obtain the observed baryon to
entropy ratio in the electroweak theory [4-9]. In particular, Nelson et al
[7,8] consider
simple extensions of the minimal standard model (MSM) to obtain
sufficient CP violation. This is provided by the complex space dependent
fermionic mass function within the bubble wall, which arises quite
generally in such models. The reflection and transmission coefficients
are then different for particles and antiparticles, leading to a separation
of some CP-odd charge. The latter is then converted to baryon asymmetry in
the broken phase by the baryon number violating process in the unbroken
phase. They consider only the propagation of the top quark through the medium
due to its large Yukawa coupling [8].

The MSM, whose CP violation was earlier thought to be too small to generate
any significant baryon asymmetry, has recently been shown by Farrar
and Shaposhnikov [9] to
have the potential to generate this asymmetry by the above mechanism,
provided one takes the quark mixing effects
at high temperature properly into account. Further they consider a direct
separation of the baryon number
by the bubble wall rather than of some other CP- odd charge.

 Although the nonminimal models were initially studied because of the
insufficiency of baryon asymmetry produced in MSM, we believe that it is
worthwhile to investigate these extensions on the cosmological front, as long
as one is looking for deviations from the MSM in the laboratory [10].

Here we reconsider the simply extended MSMs for lighter quarks with masses
up to about 5 GeV. The dispersion relation for these quarks in the
electroweak plasma is markedly different from the free one [9,11]. It is not
immediately clear how such propagations are going to affect the calculation
of the baryon asymmery.
Following Ref [9] we take the altered fermion propagation
into account to leading order and examine the generation of this
asymmetry by the so-called normal and  abnormal modes.

We avoid the details of specific models by parametrising the mass
function in a simple way and consider the direct separation of
baryon number by the wall.
We also restrict our calculation to low bubble wall velocity. The reflection
coefficients of quarks from the bubble wall are obtained in an iteration
series. We argue that our calculation, though oversimplified, does give
the order of magnitude estimate of baryon asymmetry, subject to the
uncertainties involved in the details of phase transition and other
properties of the medium.

In sec.II we review the propagation properties of quark excitations in
the electroweak plasma. We then calculate in sec.III the reflection
and transmission
currents giving rise to baryon asymmerty. These expressions are then
evaluated in sec.IV. Our concluding remarks are contained in sec.V. In the
appendix we solve the quark equation of motion within the wall in an
iteration series to fourth order.

\vspace{.8cm}

\noindent{\bf II.QUARK PROPAGATION IN HOT PLASMA}

\bigskip

Here we review the propagation properties of light quark excitations in the
electroweak plasma at high temperature. The most important effect of the
medium on the quark is that it acquires a chirally invariant effective
mass with an altered dispersion relation. Neglecting smaller contributions
due to the weak gauge boson and Higgs scalar exchanges compared to that due to
gluon exchange, the one loop self-energy leads to the same effective mass
$E_0$, for both left $(L)$- and right $(R)$ -handed quarks,
\begin {equation} E_0 = ({2\pi \alpha_s/ 3})^{1/2} T \approx {0.5 T}
\end{equation}
with $\alpha_S = .12 $ at the Z boson mass. For excitations close to $E_0$,
the effective Lagrangian incorporating the altered dispersion relation is [9]
\begin {equation}{ \cal L} = iR^\dagger (\partial_0 +{1\over 3}\sigma .\nabla
+iE_0)R+iL^\dagger (\partial_0-{1\over 3}\sigma .\nabla +iE_0)L
+mL^\dagger R +m^\star R^\dagger L \end{equation}
where we have also included the mass acquired through Higgs mechanism.

The Lagrangian (2) gives the equation of motion for the $L$ and $R$ components.
In the following we consider the one-dimensional problem where quarks
propagate along the z-axis, normal to the bubble wall. Writing
\[L=\left({\psi_1\atop \psi_2}\right), R=\left({\psi_3\atop \psi_4}\right),\]
the equations split into two independent sets. Defining
\[\Phi = \left({\psi_1 \atop \psi_3}\right), \Phi^{\prime} = \left({\psi_4
\atop
\psi_2}\right),\]
and considering solutions with positive energy $E$, they are
\begin{equation} {d\over {dz}}{\Phi} = iQ(z)\Phi ,\end{equation}
where
\begin{equation} Q(z) = 3\left(\begin{array}{cc}

E-E_0  &  m^\star\\
-m     & -(E-E_0)
\end{array}\right),\end{equation}
and a similar one for $\Phi ^{\prime }$ with $m$ replaced by its complex
conjugate.These equations refer to the fluid rest frame. Although we work
in the wall rest frame, it suffices to evaluate the reflection coefficients
in the fluid rest frame, because of our restriction to linear terms in the
bubble wall velocity in calculating the baryon asymmetry.

The planar bubble wall has a finite thickness, extending from $z=0$ to $z=z_0$.
It separates the broken phase $(z>z_0)$ from the unbroken phase $(z<0)$. The
Higgs
induced mass $m$ rises from zero in the unbroken phase through the bubble wall
to the (almost) zero temperature mass $m_0$ in the broken phase.

Requiring plane wave solutions in the broken phase, eqn.(3) yields the altered
dispersion relation mentioned above [12],
\begin{equation}E_\pm = E_0 \pm \sqrt{(p/3)^{2} +m_0^{2}}\end{equation}
In contrast to the free particle dispersion relation, where only one branch
belongs to positive energy, the presence of $E_0 ( >m_0)$ in (5) now makes
both branches accessible to a particle with positive energy. The two
branches are called normal (+) and abnormal (-) modes of propagation. Unlike
the energy $E$, the value of the variable $p$, however, does not give the
true momentum ${\bar p}_{\pm}$ of the excitation, the latter being given by
\begin{equation} {\bar p}_{\pm} = \pm{p\over{3|p|}} E_\pm \end{equation}
Also the group velocities for the two branches are given by
\begin{equation}
v_\pm = {d\over{dp}} E_\pm =  \pm {1\over3}{p\over{\sqrt{(p/3)^{2}+m_0^{2}}}}.
\end{equation}
In the unbroken phase, where the dispersion relation becomes
\begin{equation}
E_\pm = E_0 \pm k/3,
\end{equation}
each of the components $\psi_i$ satisfies an uncoupled
equation. $\psi_{1,2}$ , belonging to chirality $+1$ , has $k= \pm 3(E-E_0)$
respectively, while $\psi_{3,4}$ , belonging to chirality $-1$, has
$k = \mp 3(E-E_0)$ respectively. It might be thought that a particular
component $\psi_i$ would describe a left- or right-moving particle, depending
on whether the energy belongs to the normal $(E>E_0)$  or the abnormal
$(E<E_0)$ branch respectively. However, this is not true, as the direction
of the true momentum (and also of the group velocity ) is opposite to that of
$k$ in the abnormal mode.

Over the domain wall, m is space dependent. We may write the solution for
$\Phi (z)$ as
\begin{equation} \Phi(z) = \Omega (z) \Phi (0),    0\leq z \leq \ z_0,
\end{equation}
in terms of the $2\times 2$  unimodular matrix $\Omega (z)$. At $z=z_0$ its
elements will be denoted by
\begin{equation}
\Omega (z_0) =\left( \begin{array}{ll}
		\alpha  &\beta\\
		\beta ^\star & \alpha ^\star
		\end{array}
		\right).
\end{equation}
An iterative solution is obtained in the Appendix.

We note here the Lorentz invariant expression for the density of fermionic
excitations,
$$ n = (\exp{\beta p\cdot v} +1)^{-1} . $$
Here $\beta$ is the inverse temperature of the fluid in the frame where it is
at rest, $p^\mu$ is the energy momentum 4-vector of the excitation and
$v^\mu$ , the 4-velocity of the medium. In the wall rest frame,
$p^\mu =(E,\bar{p})$ , $v^\mu =\gamma (1,v)$ where $\gamma= 1/\sqrt{1-v^2}$
and $\bar{p}$ is the true momentum given by (6). For  $p$ along the positive
z-direction,
we thus have in this frame, $p\cdot v=E_\pm (1\mp{ v/3})$, up to linear term in
$v$. In the following we need the densities of particles moving
towards the wall. In the unbroken phase these are given by
\begin{equation}
 {n_{\pm}^u} ={1 \over{e^{\beta E_\pm (1-v/3)}+1}}
\end{equation}
for the $({\pm})$ modes respectively. In the broken phase the corresponding
quantities $ n_{\pm}^b $ are given by the same expressions with the
reversal of sign before $v$.

\vspace{.8cm}

\noindent{\bf III. REFLECTION AND TRANSMISSION CURRENTS}

\bigskip

Since baryon non-conservation through the sphaleron processes involves
the left-handed fermions and antifermions, we are interested in calculating
the left-handed baryonic currents only.

Consider first the propagation of quark excitation by the normal mode. We
send a right-handed fermion towards the domain wall from the unbroken
phase. Noting the revarsal of chirality at the wall, the incident wave (of
unit current) and the reflected wave of amplitude $r$ , say, is given by
\begin{equation}
\Phi(z) = \left({1\atop 0}\right)e^{ikz} +\left({0\atop r}\right)e^{-ikz}
,z\leq 0.
\end{equation}
On the right ( broken phase), we have only the transmitted wave of amplitude
$t$, say. Solving eqn.(3) for $\Phi$ we get
\begin{equation}
\Phi(z) = t\left({\cosh{\theta}\atop -\sinh{\theta}}\right)e^{ip(z-z_0)}
,z\geq z_0.
\end{equation}
Here $p$ satisfies eqn.(5) with the plus sign and
$$ \cosh{\theta} =\sqrt{{E-E_0 +{p/3}}\over {2p/3}}. $$

To find the unknown ampltudes we use the boundary conditions given
 by eqn.(9) for $z=z_0$,
\begin{equation}
t\left({\cosh\theta\atop -\sinh{\theta}}\right) =
\left(
\begin{array}{ll}
\alpha  &  \beta\nonumber\\
\beta^\star  & \alpha^\star \nonumber
\end{array}
\right) \left({1 \atop r}\right).
\end{equation}
The reflection coefficient is
\begin{equation}
R_+=|r|^2=1-{1\over |\alpha^\star \cosh \theta + \beta \sinh \theta|^2}.
\end{equation}
The incident current, same for particles and antiparticles, is
${1\over 3}n_+^u$, where $n_+^u$ is given by (11). Then the net contribution
to the reflected left-handed baryonic current is
\begin{equation}
\int_{3m_0}^{\infty}{dk\over 2\pi}{1\over 3}n_+^u(R_+-\bar{R}_+)
\end{equation}
Here and in the following a bar on a reflection or transmission coefficient
denotes the corresponding quantity for the antiparticle. It is obtained by
solving the same eqn.(3) with $m$ replaced by $m^\star $.

 Next we calculate the transmitted baryonic current in the unbroken phase
 due to incidence on the wall from the broken phase. On the left there is
 simply a transmitted wave of amplitude $\tilde{t}$, say,
\begin{equation}
\Phi (z) =\left({0 \atop \tilde{t}}\right)e^{-ikz}    ,  z<0 \end{equation}  On
the right we have both the incident wave (of unit current) and the reflected
wave of amplitude $\tilde{r}$, say. Solving eqn.(3) separately for the two
cases, we get
\begin{equation}
 \Phi (z) =\left({\sinh{\theta}\atop -\cosh{\theta}}\right)e^{-ip(z-z_0)}
 +\tilde{r}\left({\cosh{\theta} \atop -\sinh{\theta}}\right)e^{ip(z-z_0)}
,z\geq z_0
\end{equation}
Again using the boundary condition (9) at $z=z_0$, we have
\begin{equation}
\left({\sinh{\theta} +\tilde{r}\cosh{\theta} \atop -(\cosh{\theta} + \tilde
{r}\sinh{\theta})}\right)=
\left(
\begin{array}{ll}
\alpha  &  \beta  \\
\beta^\star  &  \alpha^\star
\end{array}\right)
\left({0 \atop \tilde{t}}\right)
\end{equation}
The transmission coefficient is then
\begin{equation}
T_{+} = |\tilde{t}|^{2} = 1- R_{+}
\end{equation}
There arises a transmitted left-handed baryonic current in the
unbroken phase given by
\begin{equation}
\int\limits _{0}^{\infty}{dp\over {2\pi}} {1\over 3} {p\over k} n_{+}^b
(T_+ -\bar{T_+})
\end{equation}

We add (15) and (20) to get the total normal mode baryonic current in the
unbroken phase due to reflection and transmission as
\begin{equation}
J_{+} =\int\limits _{3m_0}^{\infty}{dk\over{2\pi}} {1\over 3} (n_{+
}^b - n_{+}^u)(T_+ -\bar{T_+})
\end{equation}

A similar contribution to the baryonic current results from propagation by the
abnormal mode in the energy region  $E < E_0 -m_0$. It is given by
\begin{equation}
J_{-}=\int\limits _{3m_0}^{3E_0}{dk\over{2\pi}} {1\over 3}(n_{-}^b -
n_{-}^u)(T_- -\bar{T_-})
\end{equation}
where
\begin{equation}
T_{-}={1\over{|\alpha ^*\cosh{\theta}' -\beta \sinh{\theta}'|^2}}
\end{equation}
with
$$\cosh{\theta}'={\sqrt {{E_0-E+p/3}\over {2p/3}}}.$$

The total CP-violating left-handed baryonic current in the unbroken phase
generated by
reflection from and transmission through the bubble wall of the fermionic
excitations by the normal and the abnormal modes is
\begin{equation}
  J_{CP}^L= J_{+}+J_{-}
\end{equation}

The final step is to obtain the baryonic density $n_{B}$ in the broken phase
from the steady state solution to the rate equations in the two phases.
Nelson et al find numerical solution to the Boltzmann equations. We shall
follow Farrar and Shaposhnikov, who solve the diffusion equations
for small bubble wall velocity to get
\begin{equation}
n_{B} =J_{CP}^{L} f
\end{equation}
where $f$ is a given function of the diffusion coefficients for quarks
and leptons, the wall velocity and the sphaleron induced baryon number
violation rate. Their estimate for $f$ is $10^{-3}\leq f \leq 1$
in MSM, which should also be valid for its simple extensions.

\newpage

\noindent{\bf IV. ESTIMATION}

\bigskip

In the standard model with a single Higgs doublet, the expectation value of the
Higgs field is real everywhere during phase transition. Adding extra
multiplets will allow in general some of their components to acquire
space-dependent
complex values within the bubble wall. This in turn leads to a complex mass
function for the quark having Yukawa coupling to those multiplets.
 It is, in principle, derivable
from the model considered but, in practice, will depend on the (unknown) Higgs
couplings. Here we assume the simplest form for the mass function,
\begin{equation}
m(z)={m_{0}\over{z_{0}}}z+i {\delta\over z_{0}^{2}}z(z_{0}-z)
\end{equation}
within the bubble wall. The parameter $\delta$ is related to the magnitude
of CP violation in the model.

With the parametrization (27), we solve eqn.(3) by iteration.
This solution becomes a power series in three
dimensionless quantities, viz, the energy variable $y=3(E-E_0)z_0$ and the
constants $c=3m_0 z_0$ and $d=3\delta z_0$. In the Appendix we have obtained
this series up to fourth order. Inspection of the coefficients suggests
that for $y$, $c$ and $d$ less than unity, it should
represent the solution well.

The condition $y<1$ restricts $k$ to $k<1/z_0$. Now the reflection
coefficients are known to be asymptotically $\sim exp (-2\pi|k| z_0)$
which appears to set in already for $k>1/z_0$. Our approximation then
consists of truncating the upper limit of $k$ integration at $k=1/z_0$,
within which we use our solution to compute the reflection coefficients.

As long as the wall thickness $z_0\leq (15 GeV)^{-1}$ [13,14], the condition
$c<1$ would allow quarks with masses up to that of the b quark.
Note that even if this condition had allowed the t-quark mass, our result
would not apply to this quark propagation, since the dispersion relation
on which we base our work, would not be valid. Instead, the free particle
dispersion relation would be more appropriate for the t-quark, as has
been assumed in the work of Nelson et al. Finally the condition $d<1$ or
$\delta <1/{3z_0}$ is a very reasonable bound for the imaginary part of the
quark mass within the bubble wall.

It is now simple to evalulate the currents
$J_\pm$ given by eqns. (21,22). In the following we choose $z_0\simeq
(20 GeV)^{-1}$, whence the upper limit of $k$ integration is $\simeq 20 GeV$.
As the temparature of the phase transition $\sim 100 GeV$, we may expand the
density function in $v$ for small $v$ and approximate the exponentials by
unity to get
\begin{equation}
n_{\pm}^b - n_{\pm}^u \simeq -{1\over 6}\beta E_{\pm}v.
\end{equation}
The transmission coefficient may be calculated in a straightforward way
using the values of $\alpha$ and $\beta$ given in the Appendix.
We get
\begin{equation}
T_{\pm}-\bar{T}_{\pm} =\mp 4dc{\sqrt{y^2-c^2}(1/3-4y^2/45 +5c^2/42)\over
\{(1-c^2/6-c^4/168)y+c^2y^3/45+\sqrt{y^2-c^2}\}^2}
\end{equation}
Clearly for $c<1$, the higher powers of $c$ can be safely neglected.
Thus
\begin{equation}
J_\pm=\pm {\beta vdc\over 27\pi z_0}\int_c^1 dy {(E_0 \pm {y/3z_0})
\sqrt{y^2-c^2}(1-4y^2/15)\over (y+\sqrt{y^2-c^2})^2}
\end{equation}
Observe the large cancellation in the sum of $J_{+}$ and $J_{-}$ arising from
quark propagation by the normal and abnormal modes respectively.
The asymmetry current becomes
\begin{eqnarray}
J_{CP}^{L}&=&{2\beta vdc\over 81\pi z_0^2}\int_c^1 dy {y\sqrt{y^2-c^2}
(1-4y^2/15)\over (y+\sqrt{y^2-c^2})^2}\nonumber \\
&\sim& {1\over{18\pi}}\beta v \delta m_0
\end{eqnarray}
Finally noting the one dimensional entropy density $s=73\pi /{3\beta}$, the
baryon to entropy ratio is given by
\begin{equation}
n_B/s \sim 10^{-8} vf\delta m_0
\end{equation}
where $\delta$ and $m_{0}$ are expressed in GeV. Within the framework of our
calculation, it is the dynamics of the b quark propagation through the
phase transition bubbles, which gives a sizeable contribution to
$n_{B}/s$. With $v \sim 0.1$ and $\delta $ not too small, its contribution
can be comparable with the observed asymmetry, $n_{B}/s \sim 5\times 10^{-11}$.

\newpage

\noindent {\bf V. CONCLUSION}

\bigskip

We have investigated the generation of baryon asymmetry in the standard
model of the electroweak theory with more than one Higgs multiplet. Such
models generally give rise to a complex mass function for a quark within
the wall of bubbles formed during the phase transition. This constitutes
the CP violation needed for baryogenesis. We do not attempt to calculate
such a mass function, however: We paramertise its real and imaginary
parts in a simple way, making the integrals trivial to evaluate. We follow
Ref [9] to consider the direct separation of baryon number by the phase
boundary rather than separation of some other CP-odd charge to be converted
into baryon asymmetry by a separate process, as discussed in Ref [8].

The inclusion of the temperature dependent effective mass gives rise to
two modes of quark propagation in the plasma. Our calculation shows that
both modes must be considered. In fact, the net baryon asymmetry current
results after large cancellation between the baryonic currents carried
separately by the two modes.

The calculation presented here is a simple analytic one giving the order
of magnitude estimate of the baryon asymmetry. Our formula (32) neatly isolates
the model dependent parameters involved in the description of the electroweak
medium and the first order phase transition.
 These are the wall thickness $z_0$ [15], the velocity $v$ of the
medium, $\delta$ giving the magnitude of the imaginary part and $f$ related
to the plasma diffusion and sphaleron transition rate.

The calculation is limited to small bubble wall velocity in the rest
frame of the plasma, as we keep only the terms linear in $v$. Also it concerns
lighter quarks for which the dispersion relations deviate appreciably
from those of free propagation. For the t quark the free particle dispersion
relation is accurate enough and its propagation can be discussed following
the treatment of Nelson et al. Our calculation indicates that the contribution
of
 lighter quarks, like the b quark, to the baryon asymmetry can also be
substantial.

After completing the work, we learnt of several works [16,17] taking into
account the effect of damping in the quark propagation. These authors have
shown that it reduces the reflection coefficient to negligible values
making the present mechanism totally ineffective to reproduce the observed
baryon asymmetry in the minimal standard model of the electroweak theory.
However, this conclusion is not agreed by others [18], who claim that the
damping rate is irrelevant to the problem.

Even if we take the damping into account, it is easy to see that it cannot
drastically affect our result obtained in nonminimal models. In the minimal
model, the baryon asymmetry formula involves high powers of the mass matrix.
Damping effectively multiplies each such matrix by a suppression factor, making
the asymmetry negligible. However, in our formula the mass function appears
only quadratically. Furthermore, the range of values allowed for the wall
thickness $z_0$ is expected to be large in nonminimal models. For $z_0
\leq (20 GeV)^{-1}$, a simple estimate indicates that
the total suppression factor cannot be smaller than
$10^{-3}$. Inclusion of this factor still keeps our result (32) for $n_B/s$
near the observed value, given the large uncertainty in the value of $f$
mentioned at the end of sec.III.

\smallskip

One of us (S.M.) gratefully acknowledges the hospitality at the Institute for
Theoretical Physics, University of Berne, where this work was started. He also
thanks Prof. M.E.Shaposhnikov for an useful discussion.
\vspace{.8cm}

\centerline{\bf APPENDIX}

\bigskip

We solve eqn. (3) by iterating the corresponding integral equation,
\begin{equation}
\Phi (z) =\Phi (0) +i\int_{0} ^{z} dz_{1} Q(z_{1})\Phi (z_{1})
\end{equation}
For the matrix $ \Omega(z)$ in eqn. (9) the solution reads
\begin{equation}
\Omega (z) =1+i\int_{0}^{z}dz_{1} Q(z_{1}) +i^{2}\int_{0}^{z}dz_{1}
\int_{0}^{z_{1}}dz_{2}
Q(z_{1})Q(z_{2})+\cdots
\end{equation}
It is convenient to express $Q(z)$ in terms of Pauli matrices $\sigma^{m}
(m=1,2,3)$,
\[Q(z)=f_{m}(z)\sigma^{m}\]
with
\[f_{1}=-i{\rm Im}\;m, f_{2}=i{\rm Re}\;m   ,   f_{3}=E-E_{0}\]
With repeated use of
\[ \sigma^{m} \sigma^{n}=\delta^{mn}+i\epsilon ^{mnp} \sigma^{p}\]
we can reduce the products of $\sigma ^{m}$ to a linear combination of
$\sigma ^{m}$ and the unit matrix. Writing
\[\Omega(z) =\omega_{0}(z)1 +\omega_{m}(z)\sigma^{m}\]
we obtain the iterative solution for the matrix elements up to fourth order,
\begin{eqnarray}
\omega_{0}(z)  &=&  1-\int_{0}^{z}dz_{1} \int_{0}^{z_{1}}dz_{2}
{\bf f}(z_{1})\cdot {\bf f}(z_{2})\nonumber\\
&+&\int_{0}^{z}dz_{1}\int_{0}^{z_{1}}dz_{2}\int_{0}^{z_{2}} dz_{3}
\epsilon^{mnp}f_{m}(z_{1})f_{n}(z_{2})f_{p}(z_{3}) \nonumber \\
&+&\int_{0}^{z}dz_{1}\int_{0}^{z_{1}}dz_{2}\int_{0}^{z_{2}}dz_{3}\int_{0}^{z_{3}}dz_{4}
\{{\bf f}(z_{1})\cdot {\bf f}(z{_2}) {\bf f}(z_{3})\cdot {\bf f}(z_{4})
\nonumber  \\
&-& {\bf f}(z_{1})\cdot {\bf f}(z_{3}){\bf f}(z_{2})\cdot{\bf f}(z_{4})+
{\bf f}(z_{1})\cdot{\bf f}(z_{4}){\bf f}(z_{2})\cdot{\bf f}(z_{3})\}
\end{eqnarray}

\begin{eqnarray}
\omega_{m}(z) &=&  i\int_{0}^{z}dz_{1}f_{m}(z_{1})-i\int_{0}^{z}dz_{1}
\int_{0}^{z_{1}}dz_{2}\epsilon ^{mkl} f_{k}(z_{1})f_{l}(z_{2}) \nonumber \\
&-& i\int_{0}^{z}dz_{1}\int_{0}^{z_{1}}dz_{2}\int_{0}^{z_{2}}dz_{3}\{
f_{m}(z_{1})
{\bf f}(z_{2})\cdot{\bf f}(z_{3})-f_{m}(z_{2}){\bf f}(z_{1})\cdot{\bf
f}(z_{3})\nonumber\\
&+&f_{m}(z_{3}){\bf f}(z_{1})\cdot {\bf f}(z_{2})\}
+i\int_{0}^{z}dz_{1}\int_{0}^{z_{1}}dz_{2}\int_{0}^{z_2}dz_{3}\int_{0}^{z_{3}}
dz_{4}[\epsilon ^{mpq}\nonumber \\
& &\{f_{p}(z_{1})f_{q}(z_{4}){\bf f}(z_{2})\cdot{\bf f}(z_{3})
- f_{p}(z_{2})f_{q}(z_{4}) {\bf f}(z_{1})\cdot{\bf f}(z_{3}) \nonumber \\
&+& f_{p}(z_{3})f_{q}(z_{4}){\bf f}(z_{1})\cdot{\bf f}(z_{2})\}
+\epsilon ^{abc}f_{a}(z_{1})f_{b}(z_{2})f_{c}(z_{3})f_{m}(z_{4})] ,
\end{eqnarray}
Here ${\bf f}(z_1)\cdot{\bf f}(z_2)=f_m(z_1)f_m(z_2)$, for example.
With the parametrization (26) it is easy to evaluate the integrals. We actually
need $\omega _{0,m}(z_{0})$ which we simply denote by $\omega _{0,m}$.Using the
variables $y,c$ and $d$ introduced in the text and rejecting powers of $\delta
$
higher than the first, eqns.(32,33) give the following expressions
$( \omega = i\hat{\omega}_3 )$
\begin{eqnarray}
\omega _{0} &=& 1-{1\over 2}y^2 +{c^2\over 8 }+{1\over 24}y^4 - {17 c^2\over
720} y^2
+\cdots \nonumber \\
& & + d(-{c\over 180}y + \cdots ) \\
\omega _{1} &=& {c\over 6}y - {c\over 60}y^3 + {3c^3\over 560}y+\cdots
+\nonumber \\
& & + d({1\over 6} -{1\over 60}y^2+{13\over 1680}c^2 +\cdots ) \\
\omega _{2} &=& -{c\over 2}+{c\over 12}y^2 -{c^3\over 48}+ \cdots \nonumber \\
& & d({c^2\over 2520}y+ \cdots ) \\
\hat{\omega}_{3} &=& y - {1\over 6}y^3+{7c^2\over 120}y+ \cdots \nonumber \\
& & +d({c\over 30}-{c\over 315}y^2+{13 c^3\over 15120}+\cdots)
\end    {eqnarray}
The matrix elements of $\Omega(z_0)$ in eqn.(10) are given by
\[ \alpha =\omega_0+i\hat{\omega}_3,  \beta=\omega_1-i\omega_2 \]


\begin{thebibliography}{99}
\bibitem{12} In Ref.[9] nonleading contributions to the self energy are
included making the effective mass different for the left- and right-
handed fermions. This leads to four branches in the dispersion relation
instead of two considered here.
\bibitem{15} The dependence on $z_0$ in (32) is implicit through the conditions
that $y,c$ and $d$ are all less than unity.
\end{thebibliography}
\end{document}